\input stromlo

\def\etal{{\it et al.\/}~}
\def\etalk{{\it et al.},~}
\def\ie{{\it i.e.},\ }
\def\eg{{\it e.g.},\ }
\def\Msun{M$_{\odot}$}

\def\kms{\hbox{km s$^{-1}$}}

\def\Mpc{Mpc$^{-1}$} 
\def\degr{\hbox{$^\circ$}}
\def\lmean{\mathopen{<}}
\def\rmean{\mathclose{>}}
\def\MdI{$M_{\rm{d,\,IRAS}}$}
\def\MdO{$M_{\rm{d,\,opt}}$}

\title Dust and Ionized gas in Elliptical Galaxies 


\author Paul Goudfrooij$\,$^{1,2}

\shortauthor Goudfrooij

\affil @1 Space Telescope Science Institute,
       Baltimore, USA {\sl (Present Address)} \par \vskip-0.7ex 
       @2 {\it European Southern Observatory, Garching bei M\"unchen,
	Germany} 

\abstract Results from {\sl IRAS\/} and recent X-ray and optical
 surveys are reviewed to discuss the properties and 
 nature of the interstellar medium in elliptical galaxies.  \par 
 As to the dust component, there is a strong contrast with the
 situation among spiral galaxies in that masses of dust in ellipticals
 as derived from optical extinction are an order of magnitude {\it
 lower\/} than those derived from {\sl IRAS\/} data. I find that this
 dilemma can be resolved by assuming an extra, extended, {\it
 diffusely distributed component\/} of dust which is not detectable in
 optical data. \par 
 Bona-fide global correlations among ISM components
 ---and between ionized gas, dust, and global (stellar) properties of
 ellipticals--- are hard to find, which most probably reflects an 
 external origin of dust and ionized  gas in ellipticals. \par
 A strong  correlation is found, however, between the H$\alpha$+[N\II]
 luminosity and the optical luminosity within the region
 occupied by the ionized gas, which suggests hot (post-AGB
 and/or blue HB) stars within the old stellar population being a
 dominant source of ionization of the gas.  

\section Introduction 

Our understanding of the nature of the interstellar medium (ISM) in
elliptical galaxies has undergone a radical change from the consensus that
prevailed only a dozen of years ago. Recent advances in instrumental
sensitivity across the electromagnetic spectrum have revealed the
presence of a complex, diverse ISM in elliptical galaxies; 
in fact, the total mass of interstellar gas relative to that of stars
in ellipticals is similar to that in spiral galaxies, the main
difference being that the dominant gaseous component in ellipticals is
heated to the virial temperature, $\sim\,$10$^7$ K, radiating at
X-ray wavelengths. Smaller, varying quantities of H\I, CO, ionized
gas, and dust have been detected in many ellipticals as well (\eg
Bregman, Hogg \& Roberts 1992).  

Unlike the situation in spiral galaxies however, it still remains 
unclear ---and highly controversial--- what the correct description of
that ISM is, \ie what the origin and fate of the different 
components of the ISM are, and in particular whether mergers or
cooling-flows dictate the interplay between them (see Sparks,
Macchetto \& Golombek 1989; de Jong \etal 1990; Sparks 1992; 
Fabian, Canizares and B\"ohringer 1994; Goudfrooij \etal 1994b
(hereafter G+94b).


The first direct, unambiguous evidence for the common presence
of cool ISM in ellipticals was presented by Jura \etal (1987) who used
{\sl IRAS\/} ADD\-SCANs and found that $\geqsim 50$\% of nearby, bright
ellipticals were detected at 60 and 100 $\mu$m. Implied dust masses were
of order $\sim\,$10$^4 - 10^6$ \Msun\footnote*{We assume H$_0$ = 50
\kms\ \Mpc\ in this paper}. Although the finding that ellipticals
contain dust is not so surprising by itself ---after all, dust grains are
a natural product of ongoing stellar evolution---, the dust masses
observed in many giant ellipticals is in fact curiously high:\ dust grains
are destroyed within only 10$^6 \! - \! 10^7$ yr by thermal sputtering 
within the hot gas where the typical gas pressure $nT \sim
10^5$ cm$^{-3}$K (Draine \& Salpeter 1979).  

The dust component of the ISM thus represents a potentially crucial
diagnostic in establishing the true physical and evolutionary
relationships between the different components of the ISM in
ellipticals:\ the mere observed quantities of dust in ellipticals
constrain their evolutionary history. Evolutionary constraints set by dust
in ellipticals can become yet more significant when combined with
morphological and kinematical properties of the dust (cf.\ Section 2). 

The plan of this paper is as follows. I will discuss the dust
component(s) present in ellipticals in Section 2. Section 3 will deal
with correlations between the different components of the
ISM, along with possible implications of observed physical
associations of the hot, warm and cold components in selected
ellipticals. 

\section Origin and Distribution of Dust in Elliptical Galaxies

\subsection Dust found in Optical Surveys

Optical observations are essential for establishing the presence and
distribution of dust and gas in ellipticals, thanks to their high spatial
resolution. A commonly used optical technique to detect dust is by inspecting 
color-index (\eg $B\!-\!I$) images in which dust shows up as distinct,
reddened structures with a morphology different from the smooth
distribution of stellar light (\eg G+94b). 
A strong limitation of optical detection methods (as opposed to
measuring far-IR emission) is that only dust 
distributions that are sufficiently different from that of the stellar
light (\ie dust lanes, rings, or patches) can be detected. 
Moreover, detections are limited to nearly edge-on dust 
distributions, which is illustrated by the fact that no dust lanes
with inclinations $\geqsim\,$35\degr\ have been detected  (cf.\ Sadler \&
Gerhard 1985; G+94b). Thus, quoted optical detection
rates of dust represent firm lower limits by nature. Since  
an inclination range $|i| \leqsim$ 35\degr\ is equivalent to
$\sim\,$half the total solid angle on the sky, the {\it true\/}
detection rate 
should be about twice as high as the measured one (at a given
detection limit for dust absorption).
The recently measured detection rate of dust in a complete, blue
magnitude-limited sample ($B_T < 12$) of elliptical (E) galaxies in
the RSA catalog is 41\% (G+94b)
which means that {\it the vast majority of ellipticals could harbor
dust lanes and/or patches}.  

Evidently, at least part of the dust in ellipticals does not follow
the spatial distribution of the stars. This finding bears
information concerning the dynamics of this dust, \ie 
whether or not its motions are settled in the galaxy potential. 
This question is, in turn, linked to the intrinsic shape of ellipticals,
since in case of a settled dust lane, its morphology indicates 
a plane in the galaxy in which stable closed orbits are allowed (\eg Merritt
\& de Zeeuw 1983). These issues can be studied best in the inner
regions of ellipticals, in view of the short dynamical time scale 
involved, allowing a direct relation to the intrinsic shape of the 
galaxy. 
A recent analysis of {\it nuclear\/} dust properties in 64 ellipticals
imaged with HST has shown that dust lanes are
randomly oriented with respect to the major axis of the galaxy (van
Dokkum \& Franx 1995). Moreover,
the dust lane is significantly misaligned with the {\it kinematic\/}
axis of the stars for almost all galaxies in the sample of van Dokkum
\& Franx for which extensive stellar kinematics are
available\footnote*{By the way, most of the ellipticals in the
archival sample of HST images described by van Dokkum \& Franx (1995)
that turn out to be dusty were selected to be dust-free based on
ground-based observations, rendering the class of genuinely dutch-free
ellipticals very rare {\tenrm :}--)}. This means that {\it even at
these small scales}, the dust and stars are generally dynamically
decoupled, which argues for the external origin of the dust. This
conclusion is strengthened by the decoupled kinematics of stars and
gas in ellipticals with {\it large-scale\/} dust lanes (\eg Bertola
\etal 1988).  

\subsection Dust NOT found in Optical Surveys (The Diffusely
       Distributed Component of Dust)

As mentioned above, dust in ellipticals has been detected by 
optical as well as far-IR surveys. Since the optical and far-IR
surveys yielded quite similar detection rates, one is tempted to conclude that
both methods trace the same component of dust. However, this turns out
not to work in the quantitative sense:\ dust masses estimated from the optical
extinction are significantly {\it lower\/} than those estimated from
the far-IR emission (Goudfrooij \& de Jong 1995, hereafter GdJ95)\footnote{**}{The
methods used for deriving dust masses from optical extinction values
and from the {\sl IRAS\/} flux densities at 60 and 100 $\mu$m, and the
limitations involved in these methods, are detailed upon Goudfrooij \&
de Jong (1995).}.  Quantitatively,  
the average ratio $\lmean M_{\rm{d,\,IRAS}}/M_{\rm{d,\,opt}}
\rmean$ = $8.4 \pm 1.3$ for the 56 ellipticals in their sample for
which the presence of dust is revealed by both far-IR emission and optical
dust lanes or patches. 

This ``dust mass discrepancy'' among ellipticals is remarkable, since
the situation is {\it opposite\/} to that among spiral galaxies:
Recent analyses of deep multi-color imagery of dust extinction in
spiral galaxies (\eg Block \etal 1994) 
also reveal a discrepancy between dust masses derived from optical and {\sl
IRAS\/} data, {\it but in the other sense$\,$!\/}
In the case of spirals, the
discrepancy can be understood since dust temperatures of order 20 K
and lower are appropriate to spiral galaxies (\eg Greenberg \& Li
1995), whereas the {\sl IRAS\/} measurements were 
quite insensitive to ``cold'' dust at $T_{\rm d} \leq 25$ K
which radiates predominantly at wavelengths beyond 100
$\mu$m. 
Evidently, the bulk of the dust in spiral disks is too cold to emit
significantly at 60 and 100 $\mu$m, but still causes significant extinction
of optical light.  
Actually, $T_{\rm{d}} \leqsim 20\,$K is {\it also\/}
appropriate to the outer parts of ellipticals (Jura 1982; GdJ95),
making the apparent ``dust mass discrepancy'' among ellipticals even
more significant.  

GdJ95 argued that the discrepancy cannot merely
be due to orientation effects, since their Fig.\ 1 shows that the
relation between \MdI/\MdO\ and cos$\,i$ for ellipticals with regular
dust lanes is a scatter plot. This suggests that the dust in the lanes
is concentrated in dense clumps with a low volume filling factor.    
Instead, they postulate an additional, diffusely distributed component
of dust, which is therefore virtually undetectable by optical
methods. 
We note that diffusely distributed  dust is not unexpected: the
late-type stellar population of typical giant ellipticals  ($L_B =
10^{10} - 10^{11}$ L$_{\odot}$) has a substantial present-day mass
loss rate ($\sim$ 0.1 $-$ 1 \Msun\ yr$^{-1}$ of gas and dust; cf.\
Faber \&\ Gallagher 1976; Knapp, Gunn \& Wynn-Williams 1992) which can
be expected to be diffusely distributed.   
An interesting potential way to trace this diffuse component of dust is
provided by radial color gradients in ellipticals. With very 
few significant exceptions, giant ellipticals show a global
reddening toward their centers, in a sense approximately linear with
log$\,$(radius) (\eg Goudfrooij \etal 1994a, hereafter G+94a). This is
usually interpreted as gradients in stellar metallicity, as metallic
line-strength indices show a similar radial gradient (\eg Davies,
Sadler \& Peletier 1993). However, compiling all measurements
published to date on color-- and line-strength gradients within
ellipticals shows no obvious correlation (cf.\ Fig.\ 1), suggesting
that an additional process may be (partly) responsible for the color
gradients...... so what about dust$\,$? 

\figureps[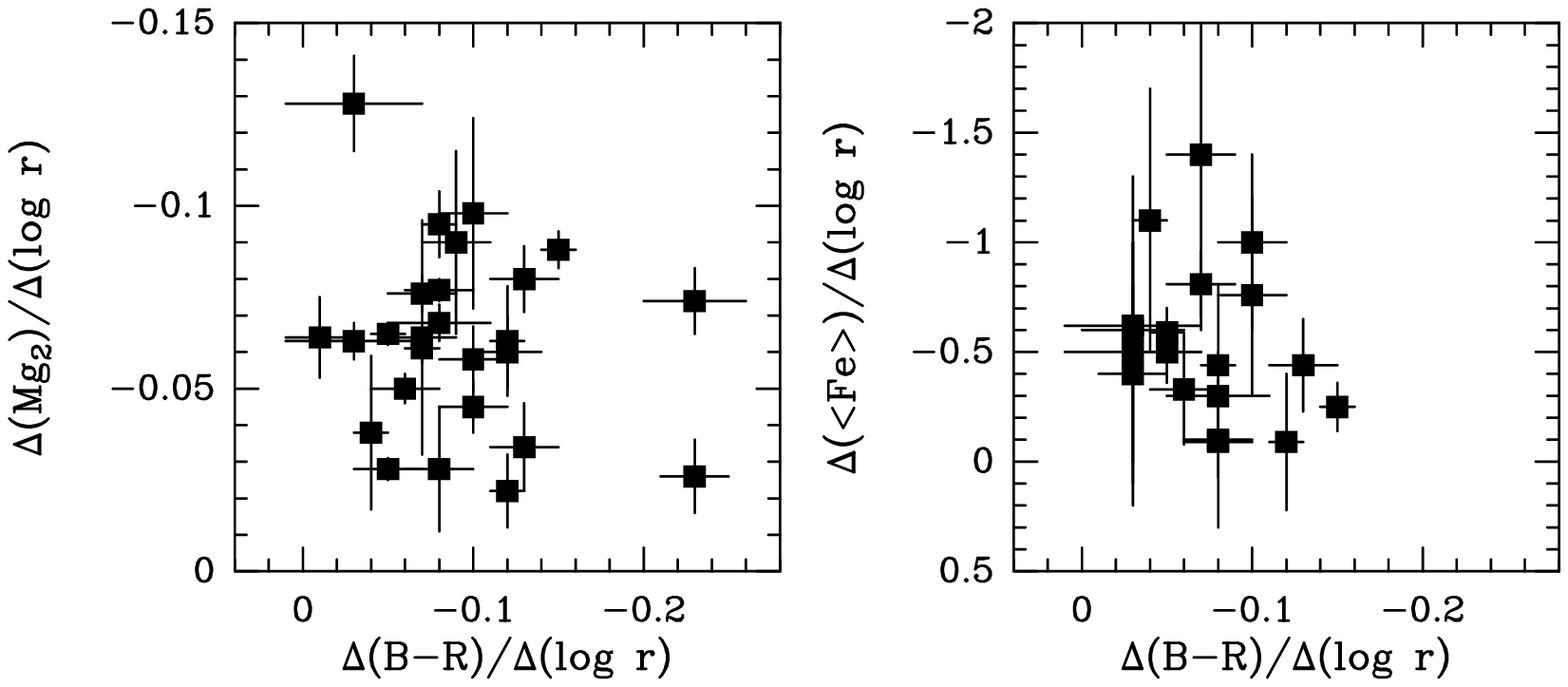,.85\hsize] 1. The relation of radial $B\!-\!R$
color gradients with radial gradients of the stellar line-strength
indices Mg$_2$ {\sl (left panel)\/} and $\lmean$Fe$\rmean$ {\sl
(right panel)} for all ellipticals for which 
both pairs of quantities have been measured to date. Data taken from Peletier
(1989), Carollo \etal (1993), Davies \etal (1993), Carollo \&
Danziger (1994), and Goudfrooij \etal (1994a). 

Indeed, recent Monte Carlo simulations of
radiation transfer within ellipticals by Witt, Thronson \& Capuano
(1992) and Wise \& Silva  (1996) have demonstrated that a diffuse
distribution of dust 
throughout ellipticals can cause significant color gradients. 
GdJ95 explored the case of an $\rho_{\rm d} \propto
r^{-1}$ density distribution for the diffuse dust component (and 
$\rho_* \propto r^{-3}$ for the stars) in ellipticals, the
rationale being that this distribution generates color gradients that
are linear with $\log(r)$, as observed (cf.\ G+94a). 
Goudfrooij \& de Jong evaluated the relation between
far-IR-to-blue luminosity ratio and color gradient as a function of
total dust optical depth (\ie the dust mass) for this case.
Their result is shown in Fig.\ 2a, which shows that color gradients in
elliptical galaxies are generally larger than can be generated by a
diffuse distribution of dust with $\rho_{\rm d} \propto
r^{-1}$. This is as expected, since color gradients should be partly
due to stellar population gradients as well. 
The distribution of the data in Fig.\ 2a is consistent with a 
``bottom-layer'' color gradient being due to differential
extinction by dust. Multivariate analysis of 
$L_{\rm IR}/L_{\rm B}$, color-- and line-strength gradients 
for a large number of ellipticals should further elucidate this matter. 

\figuretwops[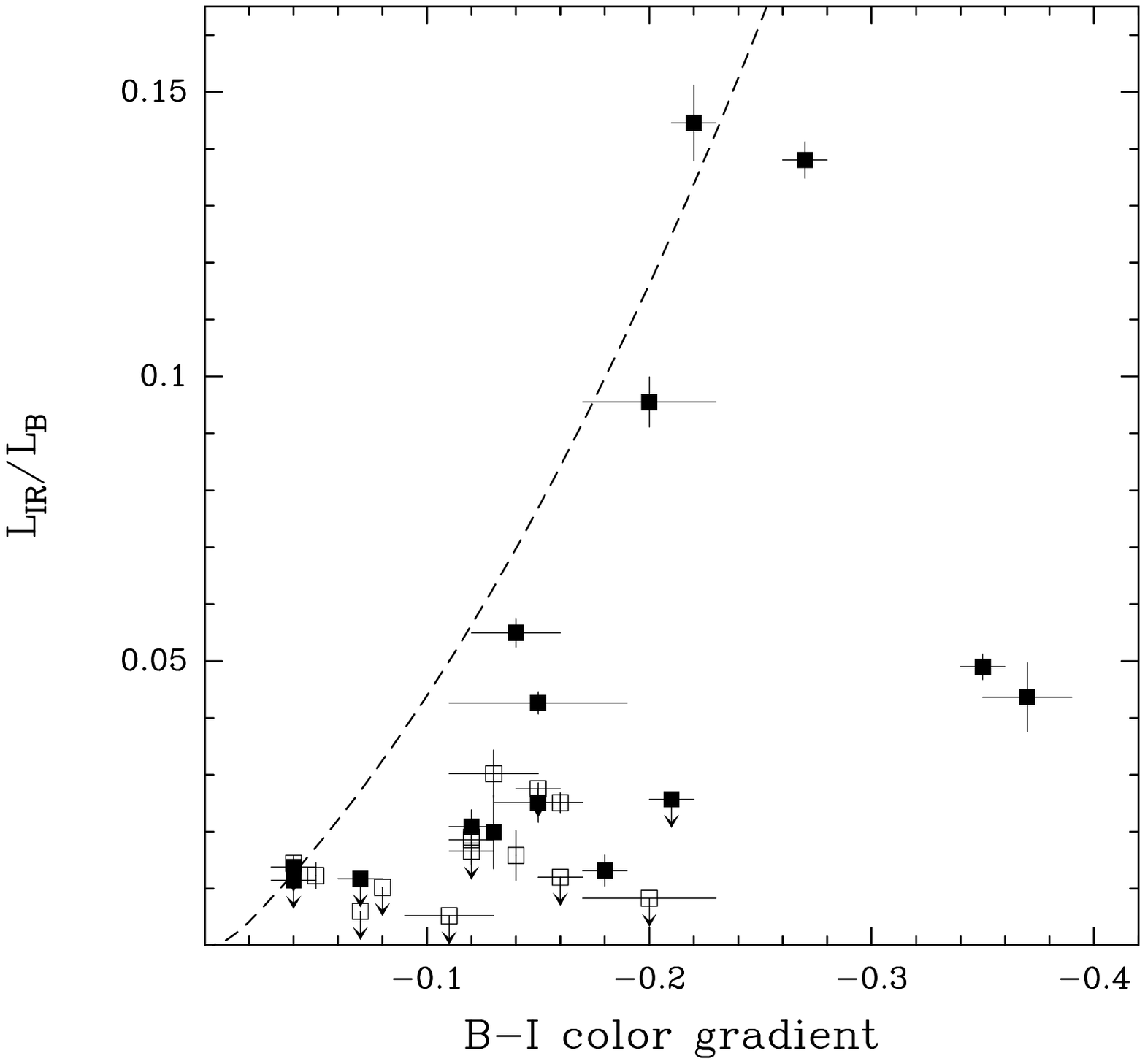,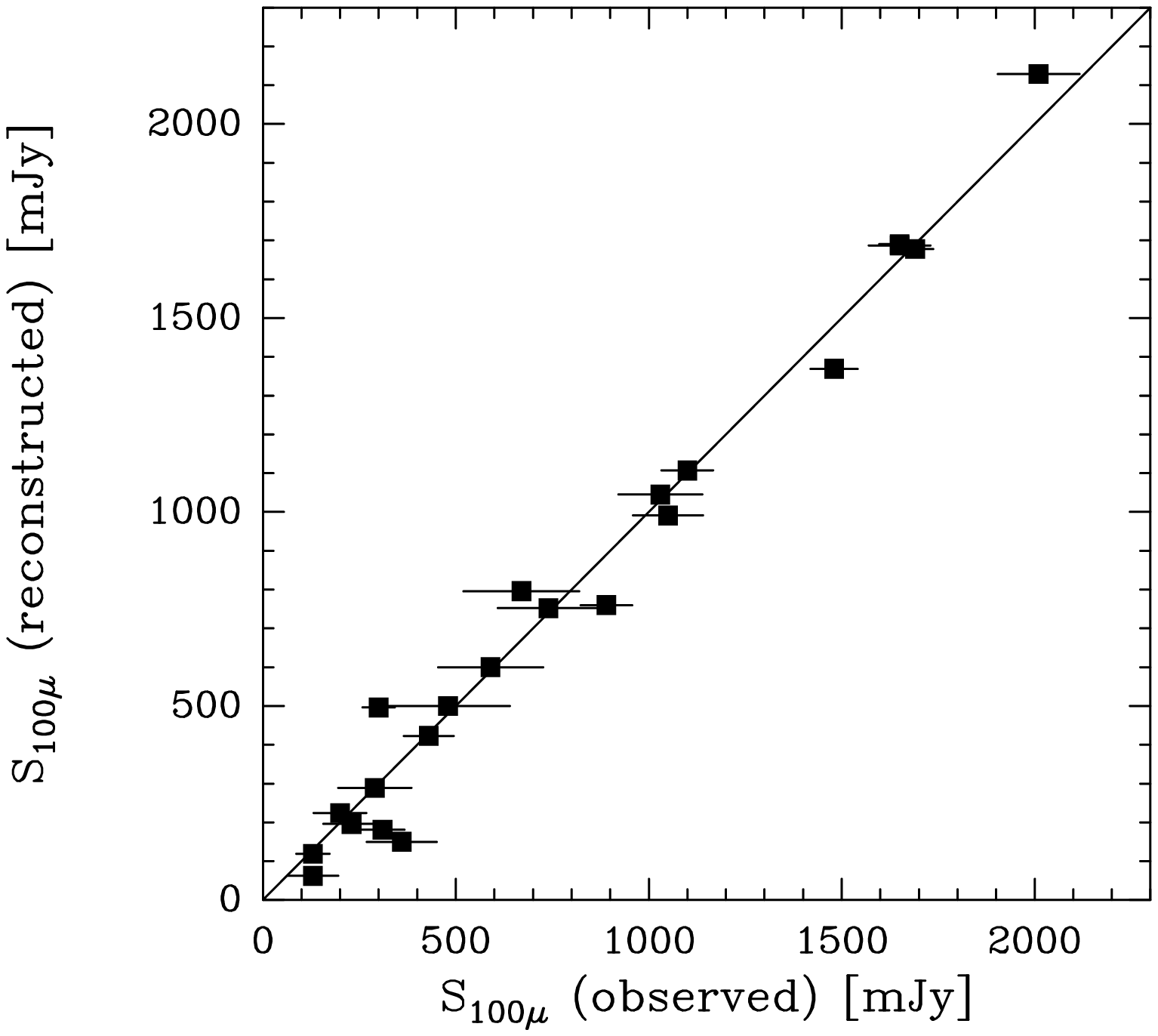,0.45\hsize] 2. {\bf (a,
left)} The relation of $L_{IR}$/$L_B$  
with $B\!-\!I$ color gradients (defined as 
$\Delta\,$($B\!-\!I$)/$\Delta\,(\log r)$) for elliptical galaxies 
in a complete, blue magnitude-limited sample. Filled squares represent
galaxies detected by {\sl IRAS\/} showing optical evidence for dust,
and open squares represent galaxies detected by IRAS without optical
evidence for dust.  Arrows pointing downwards indicate upper limits to
$L_{IR}$/$L_B$. 
The dashed line represents the color gradient expected from
differential extinction by a diffuse distribution of dust with
$\rho_{\rm d} \propto r^{-1}$ (see text). Figure adapted from
Goudfrooij \& de Jong (1995). 
{\bf ~(b, right)} 100 $\mu$m flux densities
reconstructed from calculations of dust temperatures in
ellipticals versus observed 
100 $\mu$m flux densities (and their 1$\,\sigma$ error bars). The
solid line connects loci with ``reconstructed = observed''.  Data
taken from Goudfrooij \& de Jong (1995).


Another neat feature of the $\rho_{\rm d} \propto
r^{-1}$ dust distribution is that it is also {\it energetically\/}
consistent with the available data. I don't want to go into great
detail here to illustrate this point, and refer to the discussion in
GdJ95 instead. They computed temperatures of dust
grains as a function of galactocentric radius, where the heating is
provided by stellar photons (primarily) and energetic electrons in the
hot gas (if appropriate), from which they calculated {\sl IRAS\/} 60 and 100
$\mu$m flux densities assuming $\rho_{\rm d} \propto
r^{-1}$. 
This reconstruction
reproduces the observed {\sl IRAS\/} flux densities extremely well
(cf.\ Fig.\ 2b). 

Note that so far, only dust detected by {\sl IRAS\/} has been
considered (\ie dust with $T_{\rm d} \geqsim 25\,$K); this typically
corresponds to dust within the inner few kpc.  
In reality, the diffuse component of dust in elliptical
galaxies may be expected to extend out to where the dust temperature
is lower. {\sl ISOPHOT} observations of the ``RSA sample'' of elliptical
galaxies (see G+94a) are presently being analysed, and
may reveal this cooler dust component in ellipticals. 

Apart from a distribution with $\rho_{\rm d} \propto
r^{-1}$ which supposedly accounts for the bulk of the diffusely
distributed dust, a small fraction of dust might have a steeper
distribution, being more centrally concentrated. The observational
effect of such a dust distribution has been discussed by Silva \& Wise
(1996); a hint for its presence would be the combination of a
flattened surface brightness distribution {\it and\/} a steepening of
the color gradient within the inner 10$^2$ pc of a galaxy. Future
work on inner color gradients of ellipticals will certainly resolve
this issue; 
currently available color gradients from HST WFPC2 data do {\it not\/}
show the steepening in the inner 10$^2$ pc, however (Carollo \etalk
this meeting).  

\section Correlations among various Components of the ISM

\subsection Background 

In this Section, I assemble relevant data in the literature
available to date in order to study the relationships between
the different components of the ISM in ellipticals (including
non-thermal radio emission). As to the hot and cold gaseous
components, there have been previous efforts to study their relation,
most notably:\ {\it (i)\/}  Bregman \etal (1992) who used the Roberts
\etal (1991) catalog which includes all literature data through
mid-1989, and {\it (ii)\/} Eskridge, Fabbiano \& Kim (1995) who used
the catalog of Fabbiano, Kim \& Trinchieri (1992) which comprises all
galaxies observed with the {\sl EINSTEIN\/} X-ray satellite. Since
1991, a wealth of new important data have become available, most
notably deep surveys of the ionized gas component in ellipticals. 

The existence of this ``warm'' component of the ISM in ellipticals has
been known since a long time. Early spectroscopic observations (e.g.,
Humason, Mayall \& Sandage 1956) first showed the
charateristic low-excitation nature of the emission and provided a
dectection rate of $\sim 18$\%. More recent spectroscopic surveys of
ellipticals (Caldwell 1984; Phillips \etal 1986) have pushed the
detection rate up to about 50\%. 
In order to appreciate the full spatial extent of the gaseous
emission, there have been some half a dozen recent CCD 
{\it imaging\/} surveys of the most prominent low-excitation emission lines,
H$\alpha$ and the [N\II] doublet at 6548,$\,$6583 \AA: Kim (1989)
observed 26 ellipticals and S0s detected by {\sl IRAS}, Shields (1991)
observed 46 galaxies detected at X-ray wavelengths, Trinchieri \& di
Serego Alighieri (1991; hereafter TdSA) observed 13 X-ray-emitting
ellipticals which are not at the center of clusters, G+94b observed a
optically complete sample of 56 ellipticals from the RSA catalog,
Singh \etal (1995) observed 7 X-ray-bright ellipticals and S0s, and
Macchetto \etal (1997) observed a sample of 73 ellipticals and S0s
representing a broad variety of X-ray, radio, far-IR and kinematical
properties.  
A potential concern of using H$\alpha$+[N\II] fluxes from all these
different surveys is that some surveys were much deeper than others,
which can have quite critical implications due to the fact that many
ellipticals feature extended emission at low surface
brightness. Hence, I have decided to use the flux from the deepest
survey available in case of duplicate H$\alpha$+[N\II] images of a
given object. For the galaxies in which no H$\alpha$+[N\II] emission
was detected, upper limits were calculated from the detection limit to
the emission-line surface brightness, assuming a radial extent of
1 kpc. Luminosities have been derived (or converted) according to 
$D_N\,$--$\,\sigma$ distances from Faber \etal (1989). 

Due to the page limit of this contribution, I will discuss only a few
relations involving the ISM components of ellipticals. A more
elaborate account of the statistical analysis will be published in a
forthcoming paper. 

\subsection Optical nebulosity vs.\ stellar radiation

There is a weak trend of the H$\alpha$+[N\II] luminosities with the
total B-band luminosities of the galaxies (cf.\ Fig.\ 3a), although it
is clear that there are a lot of luminous ellipticals in which no sign
of ionized gas is found; furthermore, the trend is largely due to the
{\tt [distance]}$^2$ term in the luminosities, since the trend disappears
largely in a flux-flux plot (Fig.\ 3b). 

A much more evident correlation is found between the H$\alpha$+[N\II]
luminosities and the B-band luminosity emitted within the region
occupied by the line-emitting gas\footnote*{defined as a circle 
centered on the galaxy center, and with radius equal to
$\sqrt{ab}$, where $a$ and $b$ are the semi-major and semi-minor
axis of the area occupied by the line-emitting gas, respectively.}
(Fig.\ 3c; see also Macchetto \etal 1997). This correlation does persist in a
flux-flux plot (Fig.\ 3d). Taken at face value, this correlation
suggests a stellar origin for the ionizing photons, in line with the
recent result of Binette \etal (1994) who found that post-AGB stars
within the old stellar population of ellipticals provide enough
ionizing radiation to account for the observed H$\alpha$ luminosities
and equivalent widths. Following the prescriptions of Binette {\it et
al.}, predicted H$\alpha$ luminosities have been calculated for the
current collection of galaxies. The result is L$_{{\rm H}\alpha,\,
{\rm obs}}$/L$_{{\rm H}\alpha,\,{\rm pred}} = 1.4 \pm 0.6$, which renders the
stellar origin of ionizing photons quite plausible {\it in
general}. Of course, this doesn't entirely exclude the possibility of
ionization by an active nucleus, internal shocks, or electron
conduction in hot gas in individual cases. Line-ratio studies of a
significant number of ellipticals are currently being conducted, which
can be expected to give important clues in this respect. 

\subsection Optical nebulosity vs.\ radio continuum emission

Ellipticals typically exhibit radio continuum fluxes which are
much higher than the strong correlation between
radio continuum and far-IR fluxes among spiral and S0 galaxies
would predict (e.g., Bregman \etal 
1992), and therefore thought to be primarily of nonthermal origin,
powered by an active nucleus. If the optical nebulosity in ellipticals
would also be powered primarily by active nuclei, the H$\alpha$+[N\II]
fluxes would be expected to correlate with the radio continuum
fluxes. Such a correlation is found, in fact, for powerful radio
ellipticals of the Fanaroff-Riley (FR) II class (Baum \& Heckman
1989). Radio data at 6 cm wavelength have been taken from the
literature for the ellipticals with available
H$\alpha$+[N\II] data. A plot of H$\alpha$+[N\II] luminosity vs.\ 6$\,$cm
radio power is shown in Fig.\ 3e; both quantities have been divided by
the B-band luminosity to remove the distance effect. Although
there is a weak trend for more powerful radio galaxies to have more
luminous optical nebulosity, the scatter in the plot is large. In
fact, any line-emitting galaxy with a given $L_{{\rm H}\alpha+{\rm
[N\II]}}/L_B$ ratio does not seem to ``know'' about its
radio-to-optical luminosity ratio. Thus, active nuclei do not seem to
be a significant source of ionizing photons for gas in a
``normal'' elliptical (there are no powerful FR-II type radio galaxies in
this sample).  

\figureps[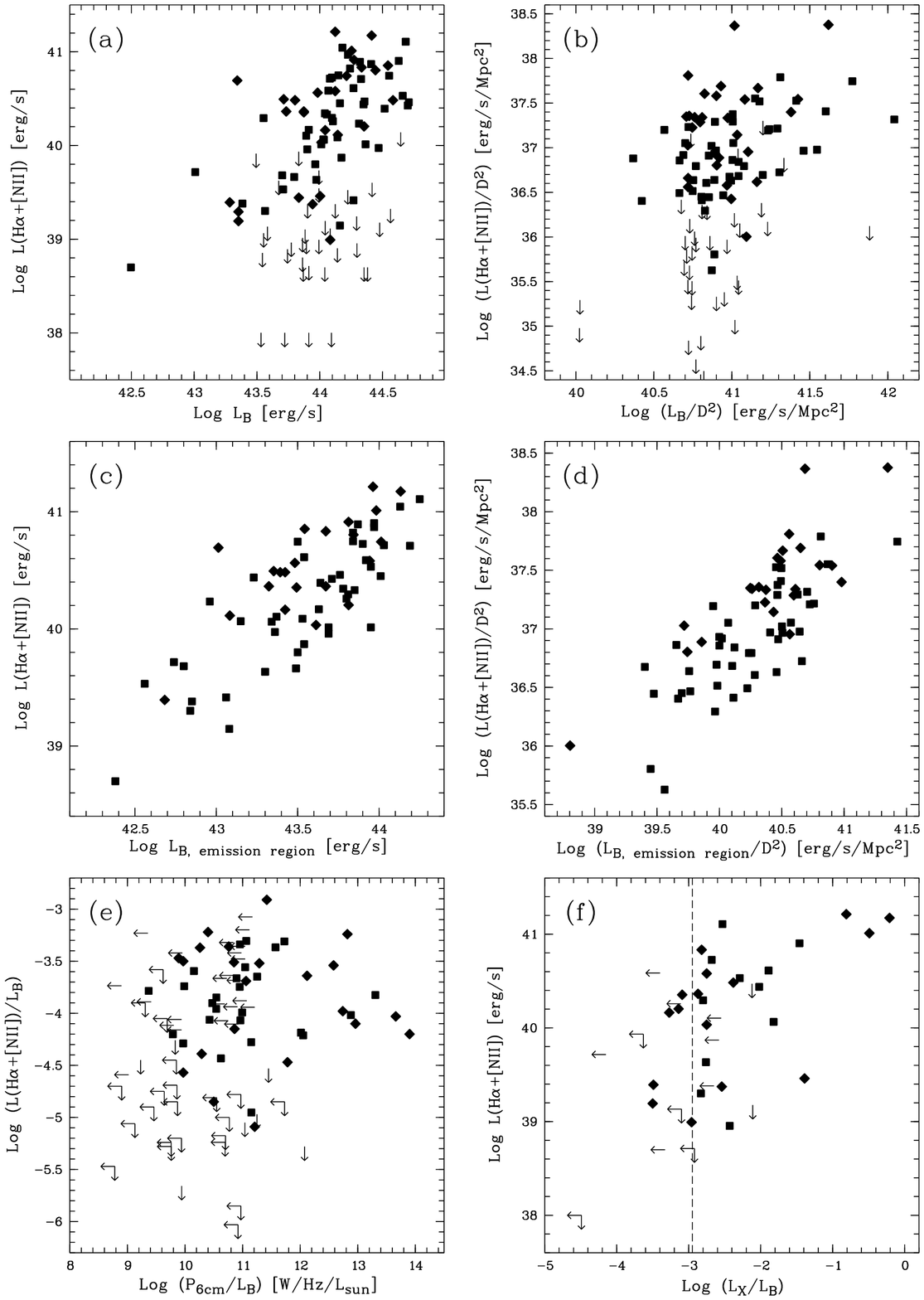,0.95\hsize] 3. Correlations among ellipticals:
{\bf (a)} H$\alpha$+[N\II] luminosity vs.\ total B-band 
luminosity; {\bf (b)} H$\alpha$+[N\II] flux vs.\ total B band flux;
{\bf (c)} H$\alpha$+[N\II] luminosity vs.\ B-band luminosity emitted
within the region occupied by the ionized gas; {\bf (d)}
H$\alpha$+[N\II] flux vs.\ flux in B band emitted within the region
occupied by the ionized gas; {\bf (e)} H$\alpha$+[N\II]-to-B-band
luminosity ratio vs.\ ratio of total radio power at 6$\,$cm over
B-band  luminosity; 
{\bf (f)} H$\alpha$+[N\II] luminosity vs.\ X-ray-to-blue luminosity
ratio. The dashed line depicts the ratio L$_{\rm X}$/L$_{\rm B}$ below
which the X-ray emission is probably not primarily due to hot gas (Kim
\etal 1992). 
{\bf Symbols:} Filled squares are data on ellipticals from
Macchetto \etal (1997) or Trinchieri \& di Serego Alighieri (1991), 
and filled lozenges are data on ellipticals from Goudfrooij \etal (1994b).  

\subsection Optical nebulosity vs.\ hot ISM

Previous studies of ellipticals have produced ambiguous conclusions
concerning possible relationships between the optical nebulosity and
the X-ray-emitting gas components.   
Shields (1991) found essentially {\it no\/} correlation, whereas
TdSA found that ---on average--- galaxies with a larger content of hot gas
also have more powerful line emission, albeit with considerable 
scatter. Both studies used an X-ray-selected sample; their different
result is possibly partly due to different observational flux
thresholds. The result of adding the new data obtained 
after 1991 is depicted in Fig.\ 3f, which shows the relation of the
H$\alpha$+[N\II] luminosities with the ratio of X-ray-to-B-band
luminosities, $L_X/L_B$. The quantity $L_X/L_B$ has been used here
rather than $L_X$ to eliminate the X-ray lumionosity due to
discrete (stellar) sources which scales linearly with B-band luminosity, so
that $L_X/L_B$ is a measure of the hot gas content; the dashed line in
Fig.\ 3f depicts the threshold in $L_X/L_B$ above which the discrete
component is supposedly negligible so that $L_X/L_B$ should scale with hot
gas content (cf.\ Kim \etal 1992).  

A glance at Fig.\ 3f reveals a correlation. However, a large
scatter is present, and there are a number of galaxies with high $L_X/L_B$
that are weak H$\alpha$+[N\II] emitters, and
several ellipticals that are bright in 
H$\alpha$+[N\II] are weak X-ray emitters. In other words, 
the nebulosity/hot gas connection is certainly not as simple as the
standard cooling flow theory (\eg Fabian \etal 1991) would predict. 
In fact, it seems advisable to study this connection on a
galaxy-by-galaxy basis rather than globally (for a whole sample). 
For example, recent individual studies of X-ray-bright ellipticals
involving optical and ROSAT HRI imaging show that the line-emitting filaments
are physically associated not only with local peaks in the X-ray
emission (as predicted in the cooling-flow theory), {\it but also with dust
lanes} (e.g., NGC$\,$4696 [Sparks \etal 1994]; NGC$\,$5846 [Goudfrooij
\& Trinchieri 1997]). This argues against the cooling-flow theory:\ 
While pressures in central regions of a cooling flow ($nT \sim 10^5 -
10^6$ cm$^{-3}\,$K) are high compared with e.g., average pressures in
the ISM of our Galaxy, they are significantly lower than those of
known sites of dust formation such as the atmospheres of red
giant stars ($nT \sim 10^{11}$ cm$^{-3}\,$K; Tielens 1990). 
At least in these cases, the alternative ``evaporation flow'' model (de Jong 
\etal 1990; Sparks 1992) ---in which the gas and dust represent ISM
brought in during a galaxy interaction--- seems to be more
appropriate. In this model, thermal interaction between the cool ISM and
the hot gas both cools the hot gas locally (thus mimicking a
cooling flow) while heating the cool ISM, giving rise to the observed
optical and far-IR emission. Future quantitative analysis of these and
new optical and X-ray data should allow us to make progress in distinguishing
between these models. 


\acknowl {\it Acknowledgments}.~~~I am very grateful to the SOC for
inviting me to participate in this great conference. Thanks are also
certainly due to Nicola Caon and Duccio Macchetto for communicating
H$\alpha$+[N\II] fluxes for early-type galaxies in their
sample to me in advance of publication.  

\references

\eightrm
\parskip=1.8pt
\baselineskip=0.78125\baselineskip
 Baum S. A., Heckman T., 1989, ApJ 336, 702 

 Block D. L., Witt A. N., Grosb{\o}l P., Stockton A., Moneti A.,
   1994, A\&A 288, 383 

 Binette L., Magris C. G., Stasin\`ska G., Bruzual A. G., 1994, A\&A
 292, 13 

 Bregman J. N., Hogg D. E., Roberts M. S., 1992, ApJ 387, 484 

 Caldwell N., 1984, PASP 96, 287

 Carollo C. M., Danziger I. J., Buson L. M., 1993, MNRAS 265, 553 

 Carollo C. M., Danziger I. J., 1994, MNRAS 270, 523 

 De Jong T., N{\o}rgaard-Nielsen H. U., Hansen L., J{\o}r\-gen\-sen 
   H. E., 1990, A\&A 232, 317

 Davies R. L., Sadler E. M., Peletier R. F., 1993, MNRAS 262, 650 

 Draine B. T., Salpeter E., 1979, ApJ 231, 77

 Eskridge P. B., Fabbiano G., Kim D.-W., 1995, ApJS 97, 141

 Faber S. M., Gallagher J. S., 1976, ApJ 204, 365
 
 Faber S. M., Wegner G., Burstein D., et al., 1989, ApJS 69, 763 

 Fabbiano G., Kim D.-W., Trinchieri G., 1992, ApJS 80, 531 

 Fabian A. C., Nulsen P. E. J., Canizares C. R., 1991, A\&AR 2, 191

 Fabian A. C., Canizares C. R., B\"ohringer H., 1994, ApJ 425, 40

 Forman W., Jones C., Tucker W., 1985, ApJ 293, 102

 Goudfrooij P., Hansen L., J{\o}rgensen H. E., et al., 1994a,
   A\&AS 104, 179  (G+94a)

 Goudfrooij P., Hansen L., J{\o}rgensen H. E., et al., 
   1994b, A\&AS 105, 341 (G+94b)


 Goudfrooij P., de Jong T., 1995, A\&A 298, 784 (GdJ95)

 Goudfrooij P., Trinchieri G., 1997, in preparation 

 Greenberg J. M., Li A., 1995, in: ``The Opacity of Spiral Disks'', 
   eds. J. I. Davies \& D. Burstein, Kluwer, Dordrecht, p. 19

 Humason M. L., Mayall N. U., Sandage A., 1956, AJ 61, 97

 Jura M., 1982, ApJ 254, 70

 Jura M., Kim D.-W., Knapp G. R., Guhathakurta P., 1987, ApJL 312, L11

 Kim D.-W., Fabbiano G., Trinchieri G., 1992, ApJ 393, 134

 Knapp G. R., Gunn J. E., Wynn-Williams C. G., 1992, ApJ 399, 76 


 Macchetto F., Pastoriza M., Caon N., et al., 
 A\&A, in press

 Merritt D., de Zeeuw P. T., 1983, ApJL 267, L19 



 Peletier R. F., 1989, Ph. D. Thesis, University of Groningen

 Phillips M. M., Jenkins C. R., Dopita M. A., Sadler E. M., Binette L.
   1986,  AJ 91, 1062

 Roberts M. S., Hogg D. E., Bregman J. E., Forman W. R., Jones C.,
 1991, ApJS 75, 751 

 Sadler E. M., Gerhard O. E., 1985, MNRAS 214, 177 



 Shields J. C., 1991, AJ 102, 1314

 Silva D. R., Wise M. W., 1996, ApJL 457, L15

 Singh K. P., Bhat P. N., Prabhu T. P., Kembhavi A. K., 1995, A\&A
 302, 658 

 Sparks W. B., 1992, ApJ 393, 66

 Sparks W. B., Macchetto F., Golombek D., 1989, ApJ 345, 153 
 
 Sparks W. B., Jedrzejewski R. I., Macchetto F., 1994, in: ``The soft
 X-ray cosmos'', eds.\ E.\ Schlegel \& R.\ Petre, AIP Press, New York,
 p.\ 389 

 Tielens A. G. G. M., 1990, in: ``Carbon in the Galaxy: Studies from
 Earth and Space'', eds.\ J. Tarter et al., NASA Conference
 Proceedings No.\ 3063, Washington, D.C., p.\ 59

 Trinchieri G., di Serego Alighieri S., 1991, AJ 101, 1647 (TdSA) 

 Van Dokkum P. G., Franx M., 1995, AJ 110, 2027 

 Wise M. W., Silva D. R., 1996, ApJ 461, 155 

 Witt A. N., Thronson H. A. jr., Capuano J. M., 1992, ApJ 393,
   611 (WTC)

 Young J. S., Schloerb F. P., Kenney D., Lord S. D., 1986, ApJ 304,
   443 

\rm
\parskip=\smallskipamount\nonfrenchspacing
\baselineskip=1.28\baselineskip

\vfill\eject

\bye